\documentclass[twocolumn,twocolumn]{IEEEtran}
\usepackage[T1]{fontenc}
\usepackage{color}
\usepackage{float}
\usepackage{mathrsfs}
\usepackage{amsmath}
\usepackage{amssymb}
\usepackage{graphicx}
\usepackage{float} 

\usepackage{CJK}
\usepackage{indentfirst}
\usepackage{amsmath}
\usepackage{cases}
\usepackage{setspace}

 \usepackage[unicode=true,
 bookmarks=true,bookmarksnumbered=true,
 bookmarksopen=true,bookmarksopenlevel=1,
 breaklinks=false,pdfborder={0 0 0},
 pdfborderstyle={},backref=false,
 colorlinks=false]
 {hyperref}

\hypersetup{pdftitle={Your Title},
 pdfauthor={Your Name},
 pdfpagelayout=OneColumn, 
 pdfnewwindow=true, 
 pdfstartview=XYZ, 
 plainpages=false}
\makeatletter
\hypersetup{draft}

\floatstyle{ruled}
\newfloat{algorithm}{tbp}{loa}
\providecommand{\algorithmname}{Algorithm}
\floatname{algorithm}{\protect\algorithmname}

\usepackage[caption=false,font=footnotesize]{subfig}
\usepackage{algorithm}
\usepackage{algorithmic}

\usepackage{multirow} 
\usepackage{amsmath} 
\usepackage{xcolor}

\allowdisplaybreaks[4]

\ifCLASSOPTIONcompsoc
\usepackage[caption=false,font=normalsize,labelfont=sf,
textfont=sf]{subfig}
\else
\usepackage[caption=false,font=footnotesize]{subfig}
\fi

\usepackage{cite}
\usepackage{bm}
\usepackage{algorithmic}
\usepackage{algorithm}
\usepackage{graphicx}
\IEEEoverridecommandlockouts

\usepackage{lettrine}

\usepackage{geometry}
\@ifundefined{showcaptionsetup}{}{%
 \PassOptionsToPackage{caption=false}{subfig}}
\usepackage{subfig}
\makeatother

\begin{document}

\title{\textcolor{black}{Routing Recovery for UAV Networks with Deliberate Attacks: A Reinforcement Learning based Approach}}

\author{
\IEEEauthorblockN{Sijie He$^{\dagger}$$^{ \parallel}$, 
Ziye Jia$^{\dagger}$$^{\parallel  }$$^{ \star  }$,
Chao Dong$^{\dagger}$, 
Wei Wang$^{\dagger}$, 
Yilu Cao$^{\dagger}$, 
Yang Yang$^{\ddagger }$, 
and Qihui Wu$^{\dagger}$\\}
\IEEEauthorblockA{$^{\dagger}$The Key Laboratory of Dynamic Cognitive System of 
 Electromagnetic Spectrum Space, Ministry of Industry and Information Technology, 
 Nanjing University of Aeronautics and Astronautics, Nanjing, Jiangsu, 210016\\
 $^{ \parallel  }$State Key Laboratory of Integrated Services Networks (Xidian University),
 Xi'an, 710071\\
 $^{\ddagger  }$Institute of Computing Technologies,
 China Academy of Railway Sciences Corporation Limited, Beijing, 10081\\
 $^{\star }$Email: jiaziye@nuaa.edu.cn}
 }

\maketitle
\pagestyle{empty} 

\thispagestyle{empty}
\begin{spacing}{1.00}
\begin{abstract}
The unmanned aerial vehicle (UAV) network is popular 
these years due to its various applications. 
In the UAV network, routing is significantly affected by the 
distributed network topology, leading to the issue that 
UAVs are vulnerable to deliberate damage. 
Hence, this paper focuses on the routing plan 
and recovery for UAV networks with attacks. 
In detail, a deliberate attack model based on the importance 
of nodes is designed to represent enemy attacks.
Then, a node importance ranking mechanism is presented, 
considering the degree of nodes and link importance. 
However, it is intractable to handle the routing problem 
by traditional methods for UAV networks, since link connections 
change with the UAV availability. 
Hence, an intelligent algorithm based on reinforcement 
learning is proposed to recover the routing path 
when UAVs are attacked.  
Simulations are conducted and numerical results verify the 
proposed mechanism performs better than other referred
methods.

\begin{IEEEkeywords}
    UAV network, trusted routing, reinforcement learning, node importance ranking mechanism.
\end{IEEEkeywords}
\end{abstract}
\end{spacing}
\newcommand{\CLASSINPUTtoptextmargin}{0.8in}

\newcommand{\CLASSINPUTbottomtextmargin}{1in}

\section{Introduction}
\begin{spacing}{1.00}
\lettrine[lines=2]{I}{n} the  next-generation mobile 
communication, unmanned aerial vehicles (UAVs) 
play important roles for full connectivity  and seamless 
coverage {\cite{9800925}}. 
In addition, UAV networks are widely leveraged 
due to the small size and flexibility in various 
applications,  such as data collection, 
real-time monitoring, search and rescue, 
surveillance, and telecommunications
 {\cite{JiaLEO,8682048,JiaHiera,9174937,you2023computation}}.
Furthermore, UAV routing is a key issue to 
support these demands.
However, the distributed topology of 
UAV networks is vulnerable to attacks, 
leading to the disconnection of communication links. 
Thus, it is significant to design an efficient  
algorithm to recover routing when the original 
path is unavailable. 

Traditional routing algorithms, 
such as Bellman-Ford and Dijkstra algorithm {\cite{9399302}},
cannot be directly applied in UAV networks,
since UAVs are characterized by high mobility, 
dynamic topology, frequent data interaction, 
and complex application environments {\cite{gupta2015survey}}. 
Moreover, the traditional algorithm cannot 
guarantee the robustness and security
for multi-UAV cooperative communications. 
Fortunately, machine learning is widely leveraged 
for routing in recent years, due to satisfactory 
performance in adapting to dynamic networks.
Reinforcement learning (RL) relies on the process 
of real-time interactions with the environment to 
update the value of 
actions performed in different states.
Meanwhile, RL-based routing algorithms 
are widely studied, since they can employ 
limited information 
to achieve satisfactory results {\cite{9738819}}.

There exist a couple of works related to routing in UAV-based scenarios. For instance, 
in {\cite{tang2022blockchain}}, Tang \emph{et al.} study the complex 
traffic offloading decision problem and then reformulate it into a 
Markov decision process (MDP) form, to reduce the total 
transmission delay in highly dynamic network environments.
Lin \emph{et al.} utilize a deep RL method to 
design an intelligent routing algorithm, in which nodes adaptively determine 
neighborhoods based on Q values {\cite{lin2022deep}}. 
In {\cite{he2020fuzzy}}, He \emph{et al.}  design a 
routing algorithm based on fuzzy logic RL, 
and simulation results show that the proposed 
algorithm has lower average hop counts
compared with the ant colony optimization algorithm.
In {\cite{9314848}}, the authors propose a deep RL-based 
adaptive and reliable routing method, considering 
the information related to routing,
such as link connections, remaining energy of nodes, and
distances, to accurately represent the network 
environment and then make appropriate routing decisions.
Liu \emph{et al.}  propose a Q-learning based multi-objective
optimization routing algorithm, which utilizes a new exploitation
and exploration mechanism to explore potential
optimal routing paths by re-estimating neighborhood relationships
during routing {\cite{liu2020qmr}}.
However, these works lack considering the issue  
that nodes are attacked and the original 
routing path is destroyed.

 Therefore, in this work, we focus on minimizing the 
 end-to-end delay of data transmission in UAV networks with deliberate attacks. 
 To clearly depict the routing process in the UAV network, 
 we present a UAV routing recovery scenario. 
Further, we design a node importance ranking mechanism 
 (NIRM), including the importance of links and 
  the node degree, to represent deliberate attacks.
 However, this problem is intractable to handle in the 
 complex UAV network. 
 Hence, we reformulate the routing issue into an MDP form and  
 then propose a RL-based intelligent algorithm.
In the end, we conduct extensive simulations to demonstrate 
the proposed method in terms of both convergence and delay.

\textcolor{black}{The rest of this paper is organized as follows. 
In Section {\ref{sec:System-Model}}, we present the system 
model and problem formulation. 
Then, the RL-based intelligent routing algorithm 
with deliberate attacks is designed 
in Section {\ref{sec:Reinforcement-learning}}. Moreover, 
Section {\ref{sec:Simulation Results}} 
provides simulation results. Finally, 
Section {\ref{sec:Conclusions}} draws conclusions.}
\end{spacing}
\begin{figure}[t]
    \centering
    \includegraphics[width=8.5cm]{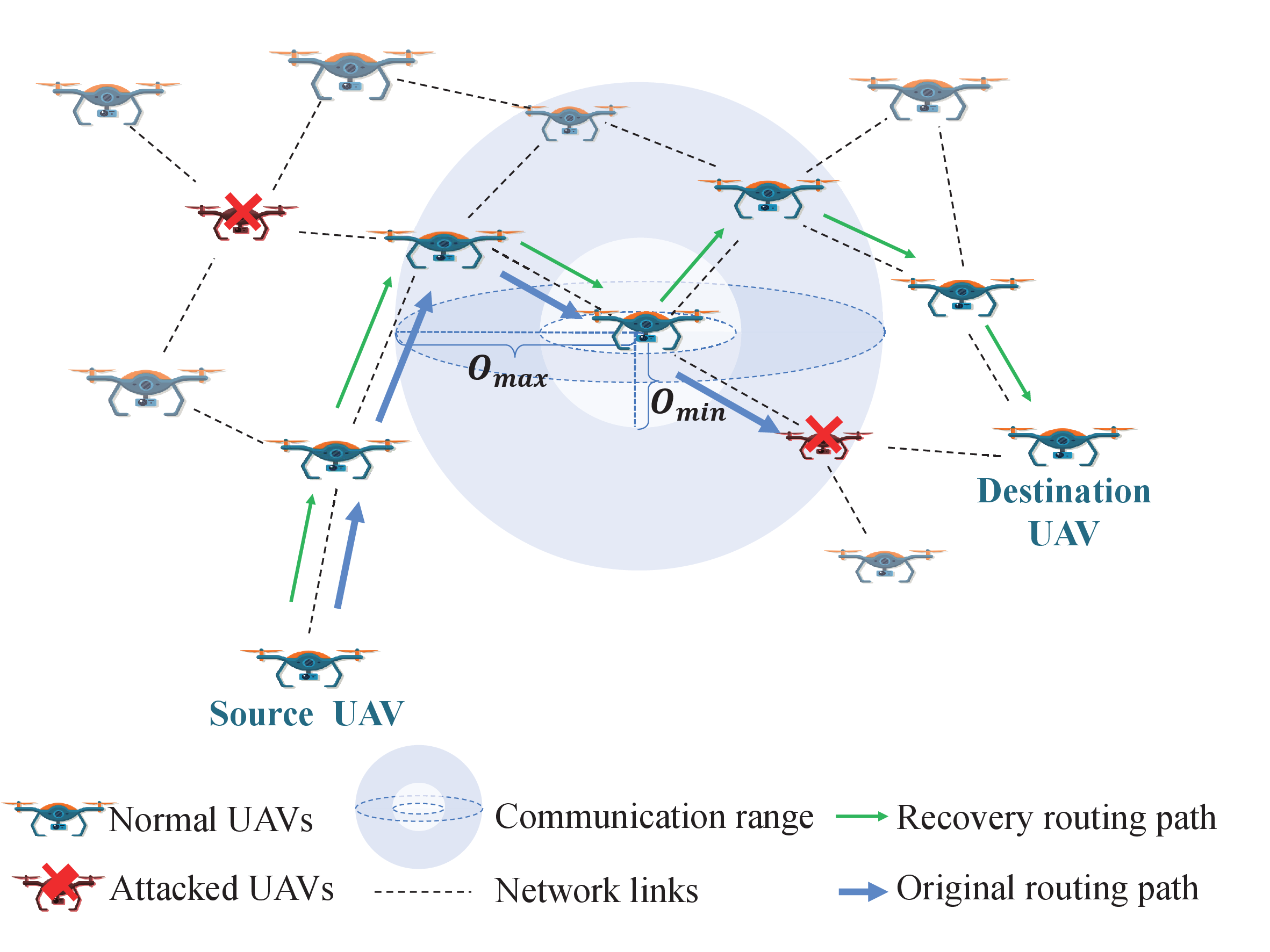}
    \textcolor{black}{\caption{\label{fig:Network_of_UAV}UAV routing recovery scenario.
    The recovery routing path is generated after UAVs in the original routing path are attacked.}}
\end{figure}

\section{System Model and Problem Formulation\label{sec:System-Model}}

\subsection{Network Model}
\begin{spacing}{1.00}
As shown in Fig. \ref{fig:Network_of_UAV}, a UAV 
network scenario is provided, in which the 
communication range of each UAV 
is limited to a hollow sphere with a maximum radius 
$O_{max}$ and a minimum radius $O_{min}$.
In detail, $O_{max}$ represents the maximum 
communicative distance and $O_{min}$ denotes the minimum 
distance between two UAVs without collisions. 
Moreover, we consider an undirected 
network $\mathcal{G}=(\mathcal{U},\mathcal{E})$ 
with $\mathcal{N}$ UAVs and $\mathcal{M}$ edges. 
$\mathcal{U}=\{{u_{i} , i=1,2,\cdots,\mathcal{N}}\}$, 
where $u_{i}$ represents the $i$-th UAV. 
 $\mathcal{E} =\{e_{i,j}, i=1,2,\cdots,\mathcal{N},j=1,2,\cdots,\mathcal{N}\}$, 
 in which $e_{i,j}\in \{{0,1}\}$ is a 
 binary variable, representing 
 if there exists an active link between 
 UAVs $u_{i}$ and $u_{j}$. 
In particular, $e_{i,j}=1$ indicates 
there exists a communication 
link between the two UAVs, 
while $e_{i,j}=0$ represents the two UAVs 
are unavailable to communicate.
Then, we define $F={\{f_{i},i=1,2,\cdots,\mathcal{N}\}}$ 
to represent the set of damaged UAVs. 
$f_{i}=1$ indicates UAV ${u_{i}}$ is attacked and 
the values of $e_{i,x}$ and $e_{x,i}$ ($x=1,2,\cdots,\mathcal{N}$) in  
adjacency matrix $\mathcal{E}$ are set as zero, 
representing that other UAVs cannot transmit data
via UAV ${u_{i}}$. 
The position of  $u_{i}$ is expressed 
by $(x_{i}, y_{i}, z_{i})$.
At the start of  routing, source node $u_{s}$ 
generates a packet of size $l_{p}$ and 
sends it to destination node 
$u_{d}\in \mathcal{U}$ via a defined routing 
path $p_{sd}=\{u_{s}\rightarrow u_{d}\}$.

\subsection{Communication Model}
According to {\cite{safwat20223d}}, the
free space path loss between two UAVs
is represented as:
\begin{equation}{\label{equ:FSPL}}
    FSPL=20 log_{10}(\frac{4\pi d g }{c}),
\end{equation}
where $d$ represents the three-dimensional 
distance between UAVs.
$g$ denotes the frequency,
and $c$ is the speed of light.
As for the communication between UAVs $u_{i}$ 
and $u_{j}$, we consider a simple 
free-space propagation model, depending on 
the frequency and distance.
$PL_{i,j}$ in $dB$ is calculated as:
\begin{equation}{\label{equ:PL}}
    PL_{i,j}=20log_{10}(d)+20log_{10}(g)-147.55.
\end{equation}

Therefore, based on ({\ref{equ:PL}}), 
the signal-to-noise ratio in the 
communication channel between 
$u_{i}$ and $u_{j}$ is:
\begin{equation}
    SNR_{i,j}=\frac{\mathcal{P}_{i,j} 10^{-\frac{PL_{i,j}}{10}}}{\sigma^{2}_{i,j}},
\end{equation}
and the max transmission rate 
is represented as:  
\begin{equation}
    \mathcal{R}_{i,j}=B_{i,j}log_{2}(1+{SNR_{i,j}}),
\end{equation}
where $B_{i,j}$, $\mathcal{P} _{i,j}$, 
and $\sigma^{2}_{i,j}$ 
denote the channel bandwidth, transmission power, and 
noise power between $u_{i}$ and $u_{j}$, respectively. 
\subsection{Delay Model}
The end-to-end delay includes the queuing 
and transmission.
Specifically, the routing delay between UAVs 
$u_{i}$ and $u_{j}$ is expressed as:
\begin{equation}{\label{equ:time}}
T_{p_{i,j}}=\frac{d_{i,j}}{c}+\frac{\mathcal{L}}{\mathcal{R}_{i,j}},
\end{equation}
where $d_{i,j}$ represents the Euclidean distance 
between  $u_{i}$ and $u_{j}$.
$c$ is the signal transmission speed.
 $\mathcal{L}$ represents the packet waiting 
 for transmission.
 Based on ({\ref{equ:time}}), we can calculate the 
 total end-to-end delay as:
\begin{equation}{\label{equ:total time}}
    {\mathcal{T}}=\underset{p_{i,j}\in{p_{sd}}}{\sum}{T_{p_{i,j}}},
\end{equation}
where ${p_{i,j}}$ denotes the one-hop path between 
 $u_{i}$ and  $u_{j}$. In addition,  
$p_{sd}$ represents a complete source-to-destination 
path from $u_{s}$ to $u_{d}$.

\subsection{Problem Formulation}
The objective is to minimize the latency of routing 
via selecting an optimal routing path
 in the UAV network with attacks. 
In detail, the problem is formulated as:
\begin{subequations}{\label{optimal}}
\begin{align}
    \mathscr{P}0:\;\underset{{p_{sd}^{b}}}{\textrm{min}}\;
    & \underset{T_{p_{i,j}}}{\sum} \underset{{U}_{\{s,\cdots,d\}}}{\sum}\mathcal{T}_{D(u_{s}\rightarrow{u_{d}},P_{sd})},\\
       \textrm{s.t.}\; 
       &T_{p_{i,j}}\leq  (T_{p_{i,j}})_{max},\\
       &O_{min}\leq  d_{i,j}\leq  O_{max},\\
       &p_{sd}^{b}\in {P_{sd}},\\
       &u_{s},u_{d}\in {U}_{\{s,\cdots,d\}},
\end{align}
\end{subequations}
where $\mathcal{T(\cdot)}$ represents the total delay 
of delivering packets in the routing path.
A routing approach is denoted as $D(u_{s}\rightarrow{u_{d}},P_{sd})$. 
In particular, $P_{sd}=\{{p^{1}_{sd},\cdots,p^{b}_{sd}}\}$ is the candidate routing path set 
and $b$ denotes the $b$-th available path. ${U}_{\{s,\cdots,d\}}$ indicates the UAV set in the routing path $p^{b}_{sd}$. 
$(T_{p_{i,j}})_{max}$ represents the tolerable maximum delay.
 $u_{s}$ and $u_{d}$ are the source and destination UAVs, respectively.

\section{Algorithm Design \label{sec:Reinforcement-learning}}
In this section, we firstly employ NIRM 
to model deliberate attacks.
Then, we reformulate the routing issue 
into an MDP form and propose a RL-based 
intelligent algorithm to make dynamic routing decisions.

\subsection{NIRM based Deliberate Attack Model}{\label{NIRM}}
In general, UAVs are characterized by high 
flexibility and complex application environments, 
and thus vulnerable to attacks.
For example, the attacker may send a large 
amount of requests to UAVs in the network, 
exhausting their network resources and making 
them unable to provide services for other UAVs, 
i.e., denial-of-service attacks.
If  middle nodes within an original routing 
path 
are attacked, another new path for routing 
should be found in limited time.

To study the impact of attacks on network performance, 
we introduce attack models. Generally, there exist 
two attack models in random and deliberate,  
respectively {\cite{wang2013vulnerability}}.
Random attack models assign the order of importance 
to nodes randomly and then launch attacks 
based on the ranking. In comparison, 
deliberate attack models rank the 
importance of nodes
according to the rules.
We consider employing the deliberate attack model 
in this work, since most attackers destroy 
the network via specific standards, 
for example, attacking critical nodes at first.  
Further, we design the deliberate
attack model based on a 
node importance ranking mechanism, 
in which UAVs with higher importance 
are attacked in priority, 
leading  to  network breakdown. 

There exist a couple of methods to evaluate node importance 
in complex networks, 
such as degree centrality (DC), closeness centrality (CC), 
and betweenness centrality (BC){\cite{liu2016evaluating}}. 
Therein, DC emphasizes the number of links connected to the node directly,
but nodes with the same degree may play different roles in the UAV network.
Moreover, BC and CC require the global information
to calculate the optimal routing path in 
large-scale UAV networks, which is intractable to obtain. 
Hence, we evaluate the importance of nodes considering both 
the node degree and link importance.
In particular, the importance of UAV $u_{i}$ is expressed as:

\begin{equation}
    L_{u_{i}} = k_{i} + \sum_{u_{j} \in \Gamma_{i}}W_{u_{i}u_{j}},
\end{equation}
where $\Gamma_{i}$ and $k_{i}$ denote the neighboring 
UAV set and the degree of $u_{i}$, respectively. 
$W_{u_{i}u_{j}}$ represents the contribution of
 $u_{i}$ to the importance of $e_{i,j}$, i.e.,
 \begin{equation}
    W_{u_{i}u_{j}}=I_{e_{i,j}}(1-\frac{k_{j} - 1}{k_{i} + k_{j} - 2}) ,
\end{equation}
where $k_{j}$ is the degree of UAV $u_{j}$, 
and $I_{e_{i,j}}$ denotes the importance 
of  $e_{i,j}$, i.e.,
\begin{equation}
    I_{e_{i,j}}={Z}\frac{2}{m+2},
\end{equation}
where $Z= (k_{i} - m - 1)(k_{j} - m - 1)$ depicts the 
connectivity ability of link $e_{i,j}$. $m$ denotes the number of 
triangles containing link $e_{i,j}$ in the UAV network topology.

\subsection{MDP based Reformulation}
During routing, packets are transmitted according 
to the designed routing path. 
The objective is to reduce the total transmission delay. 
However, it is challenging to make 
the routing decision satisfy this goal. In addition,  
the fixed routing method is inapplicable 
in the UAV network with attacks. 
Hence, it is necessary to analyze 
the current network environment to make further 
routing decisions. Thus, we reformulate
the routing issue into an MDP form,
which  is consisted of five parts: state space $S$, 
action space $A$, transition probability $P_{sa}$, 
reward function $r$, and discount factor $\gamma$. 
In addition, $\pi$ represents the policy, 
and $\tau$ is a sequence of tracks in an episode, defined as 
$\tau=\{{s_{0},a_{0},r_{1},s_{1},a_{1},r_{2},\cdots}\}$.

\subsubsection{State space}
As the input of RL models, state space $S$
indicates the environment of the agent, i.e., the UAV. 
To enable the agent aware of the environment state, 
we adopt two routing-related parameters to constitute the state space set.
The first parameter is the distance, 
which is a basic element and plays a vital role 
in most location-based routing algorithms. 
With the given location information, the Euclidean 
distance between UAVs $u_{i}$ and $u_{j}$ 
is calculated as:
\begin{equation}
    d_{i,j}=\sqrt{(x_{i}-x_{j})^{2}+(y_{i}-y_{j})^{2}+(z_{i}-z_{j})^{2}},
\end{equation}
where $(x_{i},y_{i},z_{i})$ and $(x_{j},y_{j},z_{j})$ represent 
the coordinates of $u_{i}$ and $u_{j}$, respectively.
Further, the performance of the UAV network 
is significantly affected by the business load of 
UAVs, which is related to the quality of service 
for delays. In detail, the length of queued packets is:
\begin{equation}
    \mathcal{L}=\eta\times{l_{p}},
\end{equation}
where $\eta$ represents the number of queued packets,
and $l_{p}$ denotes the size of a packet.
\subsubsection{Action space}
In UAV networks, the possible action is 
related to the neighboring UAV.
Moreover, the size of the 
 action space is equal to 
 the number of next  UAVs.
Mathematically, we represent the action space as 
 $A=\{a_{u_{i}\rightarrow u_{k}},i=1,2,\cdots,\mathcal{N},u_{k}\in \Gamma_{i}\}$,
where $a_{u_{i}\rightarrow u_{k}}$ indicates the
next hop UAV is $u_{k}$ for UAV $u_{i}$.
 \subsubsection{Transition probability}
 The transition probability is 
 represented by $P_{sa}$, 
  indicating  the probability from state 
$s \in S$ to the other state with action $a \in A(s)$.
 Since it is intractable to model accurately, 
 we design a model-free method.
 \subsubsection{Reward function}
To minimize the total end-to-end delay, 
we set $r_{t+1}$ as the instantaneous reward of performing action $a_{t}$ 
at current state $s_{t}$, i.e.,
 \begin{equation}{\label{equ:r}}
    r_{t+1}(s_{t},a_{t})=(-100\times T_{p_{i,j}})\mathcal{H}_{k},      
\end{equation}
where $T_{p_{i,j}}$ is the delay of ({\ref{equ:time}}).
$\mathcal{H}_{k}$ is defined as the flag 
for the end of routing processes, i.e.,
    \begin{equation}
        \mathcal{H}_{k}=\left\{
        \begin{aligned}
        0 & , & u_{k}=u_{d}, \\
        1 & , & u_{k}\neq u_{d},
        \end{aligned}
        \right.
    \end{equation}   
where $u_{d}$ is the destination UAV of 
routing path $p_{sd}$.
\subsubsection{Discount factor}
The discount factor is indicated as $\gamma \in (0,1)$, 
to calculate the cumulative reward. 
It denotes the impact of the future reward.  
A larger $\gamma$ indicates the decision 
focuses on the long-term reward.

The policy $\pi$ leads the agent to select an
action $a\in A(s)$ under state $s$.
According to the MDP model and a specific policy $\pi$, 
we can obtain a routing path. 
Hence, the objective function (\ref{optimal}) 
is transformed to find the optimal policy $\pi^{\star}$, 
which minimizes the total end-to-end delay.
Since the MDP model is the abstract of the environment in RL, 
we design a RL-based intelligent approach to tackle the issue.

\subsection{Intelligent Routing Algorithm}

Temporal difference (TD) via combining Monte Carlo (MC) with 
dynamic programming, is a fundamental method applied 
in model-free RL with incomplete traces {\cite{8742790,9362284}}.
In addition, it learns the action value functions of successor states 
to approximate the  current state. 
Sarsa is an on-policy TD algorithm, i.e., 
it utilize the same policy for sampling and evaluating.
In Sarsa, the optimal policy $\pi^{*}$ is gained in the process 
of continuous interactions with the environment. 
The goal is to find  $\pi^{*}$ to
maximize  cumulative  expected reward $R$, i.e.,
\begin{equation}
    \pi^{*}=\underset{\pi}{\text{arg max }} R.
\end{equation}

According to the policy optimization theorem, the optimal policy $\pi^{*}$ is equal to  
the optimal action value function $ Q^{*}$. 
Therefore, we can obtain $\pi^{*}$ via finding $Q^{*}$.
Further, if $Q^{*}(s,a)$ is known,
the optimal policy $\pi^{*}(a|s)$ corresponds to perform
action $a$ under state $s$, i.e.,
\begin{equation}
    \pi^{*}(a|s)=\left\{
        \begin{aligned}
            1 & , \text{ } a=\underset{a\in A(s)}{\text{arg max } }Q^{*}(s,a), \\
            0 & , \text{ otherwise},
        \end{aligned}
    \right.
\end{equation}
where $Q(s,a)$ is the value of selecting 
action $a$ under given state $s$, 
represented by the expected future reward:
 \begin{equation}
       Q(s,a)=E{\left[r_{t+1}+\gamma{{Q(s_{t+1},a_{t+1})}}|s_{t}=s,a_{t}=a\right]},
 \end{equation}
 where  $a_{t}$ and $s_{t}$ represent the current action and state, 
 while $s_{t+1}$ and $a_{t+1}$ are the next state and action, respectively.

In addition, in the learning progress, the following formula is 
represented to update $Q(s_{t},a_{t})$ at time step $t$, i.e.,
\begin{equation}
   Q(s_{t},a_{t})\leftarrow{Q(s_{t},a_{t})+\alpha  \delta_{t}},
\end{equation}
where $\alpha$ represents the learning rate, 
and action value function error $\delta_{t}$ is denoted as:
\begin{equation}{\label{equ:delta}}
    \delta_{t}=r_{t+1}+\gamma{Q(s_{t+1},a_{t+1})}-Q(s_{t},a_{t}),
 \end{equation}
where  $Q(s_{t+1},a_{t+1})$ represents 
the action value function at time step $t+1$. 
The term $r_{t+1}+\gamma{Q(s_{t+1},a_{t+1})}$ is the action value function target.

However, the Sarsa method updates $Q(s_{t},a_{t})$ 
only according to the action value function
of time step $t+1$, while the MC method 
requires reaching the final 
state of an episode to update. 
By partly leveraging these two methods, 
we design the $n$-step Sarsa method, i.e., 
Sarsa($\lambda$), where $n$ is a hyperparameter. 

There exist two methods, including the forward and backward Sarsa($\lambda$).
In particular, the forward method gives the weight 
$(1-\lambda)\lambda^{n-1}$ for each step reward. 
It requires a lot of subsequent state values, 
and intractable to implement. 
In contrast, we adopt the backward method which 
introduces the eligibility trace to each state, to 
indicate the effect on subsequent states. 
In detail, it proportionally assigns $\delta_{t}$  
to other action value functions as the update basis.
In {\cite{9362284}},
 the eligibility trace provides the short-term memory of 
 traces. Moreover, the experienced state 
 is no longer immediately deleted, 
 and some information is preserved. 
The accumulating eligibility trace is updated 
according to the following rule:
 \begin{equation}{\label{equ:ET}}
    \left\{ 
        \begin{aligned}
            &E_{0}(s,a)=0,\\
            &E_{t}(s,a)=\gamma\lambda E_{t-1}(s,a)+1(s_{t}=s,a_{t}=a),
        \end{aligned}
    \right.
\end{equation} 
where $\lambda \in (0,1)$ represents the degradation parameter.
$1(s_{t}=s,  a_{t}=a)$ is a judgment expression, 
in which the value is 1 when $s_{t}=s \text{ and }a_{t}=a$, 
and otherwise
the value is 0.
Correspondingly, $Q(s_{t}, a_{t})$ is updated as:
\begin{equation}{\label{equ:q_new}}
    Q(s_{t},a_{t})\leftarrow{Q(s_{t},a_{t})+\alpha\delta_{t}E_{t}(s,a)}.
\end{equation}

Moreover, to balance the exploration and exploitation 
 in continuously interactive processes and avoid 
 obtaining a suboptimal solution, the $\epsilon$-greedy policy is generally 
 designed to select actions, i.e.,
\begin{equation}{\label{equ:a}}
    a=\left\{
    \begin{aligned}
        &\text{random action,} \text{ probability $\epsilon$},\\
        &\text{action with $Q_{max}$,} \text{ probability $1-\epsilon$},
    \end{aligned}
    \right.
\end{equation}   
where $\epsilon$ represents the probability of exploring actions, and $Q_{max}$ denotes 
the maximum $Q$ value.

The proposed UAV attack-based Sarsa$(\lambda)$ 
algorithm is detailed in Algorithm 
{\ref{Alg:Sarsa(lambda)}}. Firstly, we assume that
there exist $N$ episodes. At the beginning of each episode, 
$Q(s,a)$ is set arbitrarily, and 
$E(s,a)$ is set as zero (line {\ref{initialize}}). 
Then, we determine whether UAV $u_{i}$ is attacked based on 
the value of $f_{i}$. If $u_{i}$ is damaged, 
the corresponding values in adjacency matrix $\mathcal{E}$
turn to zero (line {\ref{benginfor}}-{\ref{endfor}}).
Furthermore, the state  is initialized by  source UAV 
$u_{s}$, and the action is set as the next hop of  
$u_{s}$ (line {\ref{i_a}}). 
Then, the reward and next state are obtained after 
performing action $a$ under current state $s$ in the environment (line {\ref{R}}).
In addition, the next action is refreshed according to the designed $\epsilon$-greedy 
policy in (\ref{equ:a}) (line {\ref{a}}).
$\delta$, $E(s,a)$, and $Q(s,a)$ are updated 
according to (\ref{equ:delta})-(\ref{equ:q_new}), respectively.
Furthermore, the new state and action are refreshed 
(line {\ref{s_a}}). Finally, this episode terminates 
when the next hop is destination UAV $u_{d}$.

\begin{algorithm}[t]
    \caption{UAV Attack-based Sarsa$(\lambda)$ Algorithm \label{Alg:Sarsa(lambda)}}
\begin{algorithmic}[1]

\REQUIRE {$N$, ${\alpha}$, ${\gamma}$, $F$, and  $\mathcal{E}$}.

\ENSURE Optimal policy ${\pi^{*}}$.
\STATE{\label{initialize}} 
\textit{Initialization:} set $Q(s,a)$ arbitrarily 
and $E(s,a)=0$ for each $s \in S$ and $a \in A (s)$.
\REPEAT
\FOR{each $f_{i}\in F$}{\label{benginfor}}

\IF{$f_{i}==1$} 

\STATE {\label{zero}} The values of $e_{i,x}$ and $e_{x,i}$ ($x=1,2,\cdots,\mathcal{N}$) in  
adjacency matrix $\mathcal{E}$ are set as zero.

\ENDIF

\ENDFOR{\label{endfor}}

\STATE {\label{i_a}} Initialize state $s$ as source UAV $u_{s}$, and select action $a$ 
according to the designed $\epsilon$-greedy policy.
\REPEAT

\STATE {\label{R}}$r,s^{\prime}\leftarrow{ \text{state-action }  }(s,a)$.
\STATE {\label{a}}$a^{\prime}\leftarrow\epsilon \text{-greedy policy }  (Q,s^{\prime})$.
\STATE $\delta\leftarrow{r+\gamma{Q(s^{\prime},a^{\prime})-Q(s,a)}}$.
\STATE  $E(s,a)\leftarrow E(s,a)+1$.

\FOR{all $s\in{S}$, $a\in {A}(s)$}

\STATE $Q(s,a)\leftarrow Q(s,a)+\alpha\delta E(s,a)$.
\STATE $E(s,a)\leftarrow \gamma\lambda E(s,a)$.

\ENDFOR

\STATE \label{s_a} $s\leftarrow s^{\prime}, \text{ and }  a\leftarrow a^{\prime}$.

\UNTIL $s$ is the destination UAV.
\UNTIL all episodes are executed.

\end{algorithmic}

\end{algorithm}

\begin{figure}[t]
    \centering
    \subfloat{\includegraphics[width=8.8cm]{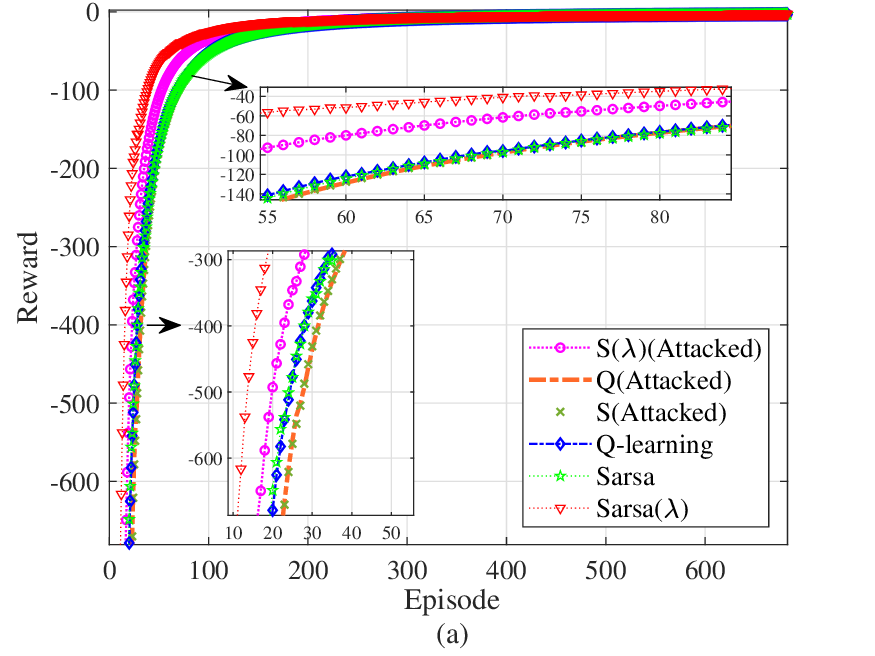}}\hspace{5pt}
	\subfloat{\includegraphics[width=8.8cm]{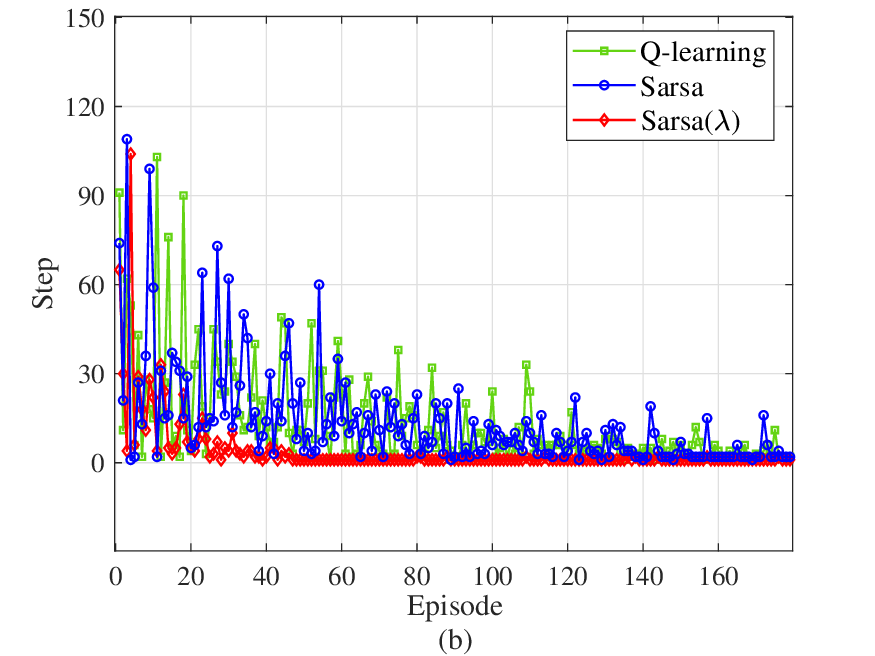}}\hspace{5pt}
    \caption{\label{fig:Reward_steps}(a) Reward \emph{v.s.} Episode. (b) Step \emph{v.s.} Episode.}\hspace{5pt}
\end{figure}

\section{Simulation Results \label{sec:Simulation Results}}
\subsection{Experiment Setup}
We employ MATLAB as the platform to train 
the proposed RL-based intelligent routing mechanism. 
In detail, the UAV network simulation area is set as 
1km$\times$1km, and the height 
of UAVs changes from 130m to 140m. 
In addition, the maximum and minimum transmission 
ranges of each UAV are initialized as 500m and 30m, respectively. 
Moreover, the source and destination UAVs 
are randomly predetermined.
Then, the length of queued packets  is generated by 
following a uniform distribution.
Other parameters are set as:
$\gamma=$ 0.9, $\alpha=$ 0.01, $\lambda=$ 0.9, 
$\mathcal{P}_{i,j}=$ 40W, $\epsilon=$ 0.001, $g=$ 2.4GHz,
${B}_{i,j}=$ 4Mhz, $\sigma^{2}_{i,j}=$ 4e-13W,  
 $\eta \in $ [1,5], and $l_{p}=$ 512Bytes
 {\cite{tang2022blockchain,lin2022deep,JiaLEO}}.
To demonstrate the performance of Algorithm {\ref{Alg:Sarsa(lambda)}}, 
two related methods are compared, i.e.,
\subsubsection{Sarsa}
Sarsa($\lambda$) algorithm is an extension of  the
Sarsa method. In addition, Sarsa($\lambda$) introduces the eligibility trace,   
which increases the state and action weights of UAVs 
closest to the destination, speeding up the  convergence of algorithm. 

\subsubsection{Q-learning}
Different from Sarsa and Sarsa($\lambda$), 
Q-learning is an off-policy algorithm. 
It leverages $\epsilon$-greedy methods to 
select $Q(s_{t}, a_{t})$ 
and employs a maximum value algorithm 
to update $Q(s_{t+1}, a_{t+1})$.

\subsection{Simulation Results}

\begin{figure}[t]
    \centering
    \includegraphics[width=8.8cm]{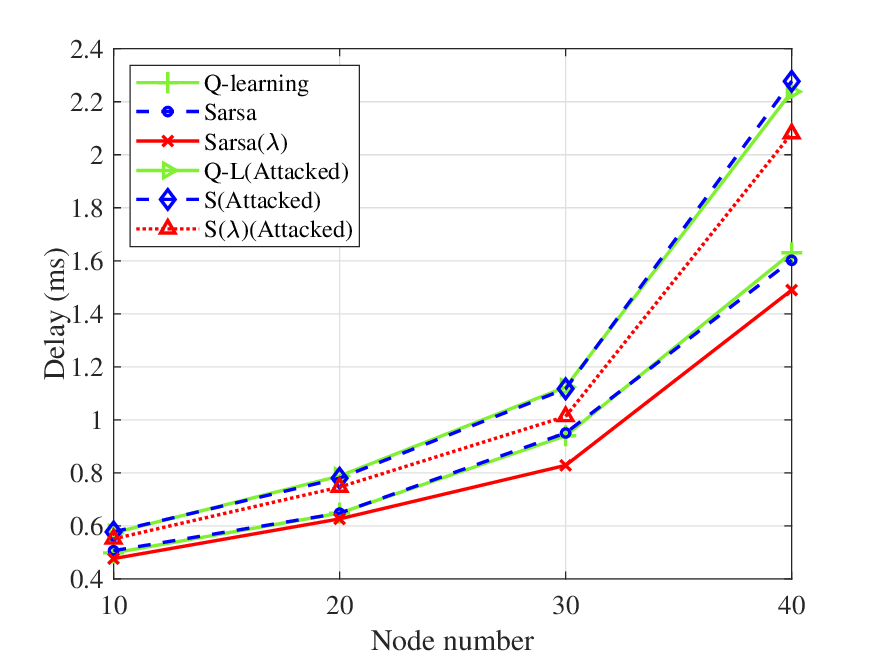}
    \textcolor{black}{\caption{\label{fig:delay}Delay \emph{v.s.} Node number.}}
\end{figure}

\begin{figure}[t]
    \centering
    \includegraphics[width=8.8cm]{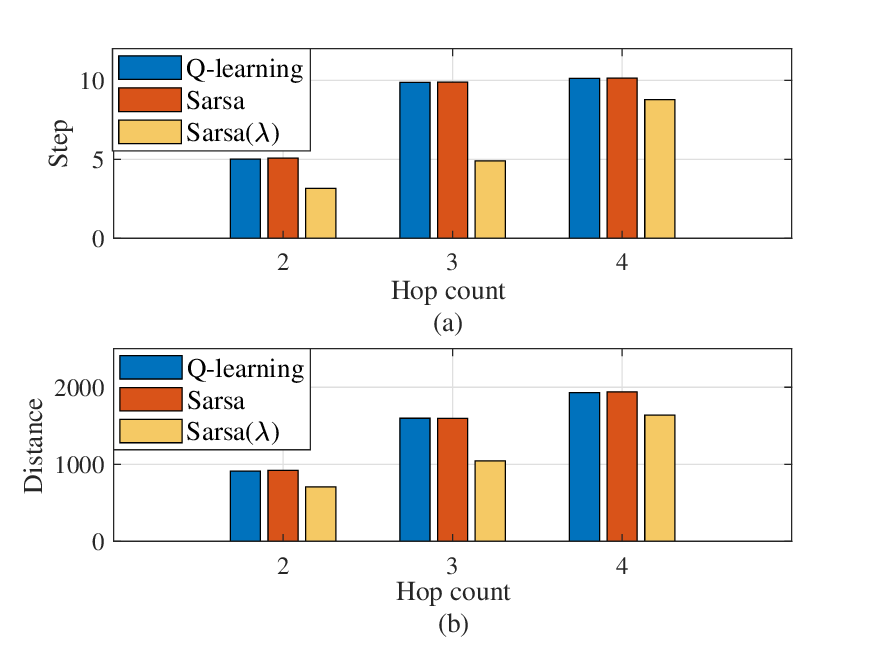}
    \textcolor{black}{\caption{\label{fig:steps_distance} 
    (a) Step \emph{v.s.} Hop count. (b) Distance \emph{v.s.} Hop count.}}
\end{figure}

To evaluate the performance of the proposed method, 
simulation results are compared in Fig. {\ref{fig:Reward_steps}-\ref{fig:steps_distance}}. 
In Fig. {\ref{fig:Reward_steps}}, 
the convergence of Sarsa($\lambda$) is depicted by 
the value of the reward  in (a) 
and the curve fluctuation of step counts in (b) 
with increasing episodes ($\mathcal{N}=20$). 
In detail, as shown in Fig. {\ref{fig:Reward_steps}}(a), 
it is obvious that the reward eventually 
converges to the maximum value, 
i.e., the minimum delay. Moreover, 
when the UAV network is attacked, 
the reward decreases, 
but Sarsa($\lambda$) still obtains the higher 
reward than the other two methods. 
In other words, Sarsa($\lambda$) leverages 
fewer episodes to find the optimal routing path. 
In addition, the reward is negative 
due to designed (\ref{equ:r}). 
Meanwhile, from Fig. {\ref{fig:Reward_steps}}(b), 
it is observed that Sarsa($\lambda$)  
has a much smaller range of curve 
fluctuations in step counts, 
since the eligibility trace is introduced.
Hence, the faster convergences of the reward and 
step counts reveal the superiority and effectiveness 
of Algorithm {\ref{Alg:Sarsa(lambda)}}.
In addition, from Fig. {\ref{fig:delay}}, the delay 
of routing has an 
exponential growth as the number of UAVs increases.  
Specifically, Sarsa($\lambda$) costs 
less time in both original and recovery 
routing paths, due to the better performance of its convergence compared with the other two methods. 
Similarly, in Fig. {\ref{fig:steps_distance}}, the numbers of 
average steps and distances in 
different hops are shown, where Sarsa($\lambda$) 
needs fewer average steps and distances for routing.
In short, simulation results illustrate 
that the proposed algorithm recovers 
the routing path with shorter delays. 

\section{Conclusions\label{sec:Conclusions}}
In this paper, we study the issue of routing  
 in the UAV network with deliberate attacks. 
The optimization objective is to minimize the total end-to-end 
delay between source and destination UAVs by selecting 
the optimal routing path. Then, we represent attacks 
on UAVs via the deliberate attack model based on NIRM. 
Further, the problem is reformulated into an MDP form, 
and we design the RL-based intelligent algorithm. 
Simulations are conducted and results verify that 
the proposed algorithm can recover the routing 
path in less time. 
\end{spacing}
\textcolor{black}
 {
     \bibliographystyle{IEEEtran}
     \bibliography{ref2}
 }
\end{document}